# MULTI-OBJECTIVE GEOMETRIC PROGRAMMING PROBLEM BEING COST COEFFICIENTS AS CONTINUOUS FUNCTION WITH WEIGHTED MEAN METHOD

A. K. Ojha and A.K. Das

**Abstract**- Geometric programming problems occur frequently in engineering design and management. In multi-objective optimization, the trade-off information between different objective functions is probably the most important piece of information in a solution process to reach the most preferred solution . In this paper we have discussed the basic concepts and principles of multiple objective optimization problems and developed a solution procedure to solve this optimization problem where the cost coefficients are continuous functions using weighted method to obtain the non-inferior solutions.

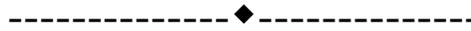

## 1. INTRODUCTION

Geometric programming (GP) derives its name from its intimate connection with geometrical concepts because the method based on geometric inequality and their properties that relate sums and products of positive numbers. Its attractive structural properties as well as its elegant theoretical basis have led to a number of interesting applications and the development of numerous useful results. The integrated circuit design, engineering design project management and inventory management are examples. Geometric programming problems (GPPs) are smooth non-linear programs in which the objective and each constraint function is a posynomials i.e. a linear combination of terms with each term a product of variables raised to real powers and each constraint function must be $\leq$ 1.The decision variables $x_j$ are restricted to be positive, to ensure that terms involving variables raised to fractional powers are defined.

If all the linear combination coefficients are positive, the functions are called posynomials and the problem is easily transformed to a convex program in new variables $y_j = lnx_j$.

- *Dr. A.K. Ojha, School of Basic Sciences, I.I.T., Bhubaneswar, Orissa, Pin- 751013, India.*
- *A.K. Das, Department of Mathematics, Bhadrak College, Bhadrak, Orissa, Pin- 756100, India.*

Otherwise the general posynomial problem is non-convex. Most of these GP applications are posynomial type with zero or few degrees of difficulty. The degree of difficulty is defined as the number of terms minus the number of variables minus one, and is equal to the dimension of the dual problem. When the degree of difficulty is zero, there is a unique dual feasible solution. If the degree of difficulty is positive, then the dual feasible region must be searched to maximize the dual objective, while if the degree of difficulty is negative, the dual constraints may be inconsistent. For detailed discussions of various algorithms and computational aspects for both posynomial and signomial GP refers to Beightler [2], Duffin [7], Ecker [8] and Phillips [15]. Generally, an engineering design problem has multiple objective functions that are usually non-commensurable and in conflict. An ideal solution is that which is optimal with respect to all objectives in general .Trade-offs must often be made between different objective functions. This concern has led to a steady advancement in the research of multi-objective optimization during the last three decades. Biswal [4] has studied the optimal compromise solution of multi-objective programming problem by using fuzzy programming technique [22, 23]. In a recent paper, Islam and Ray [9] find the pareto optimal solution by considering a multi-objective entropy transportation problem with an additional restriction with generalized fuzzy number cost.

In this paper, we have developed the method to find the compromise optimal solution of certain multi-objective geometric programming problems where the cost coefficients are continuous functions by using weighting method. First of all, the multiple objective functions transformed to a single objective by considering it as the linear combination of the multiple objectives along with suitable constants called weights. By changing the weights, the most compromise optimal solution has been arrived by using GP techniques.

The organization of the paper is as follows: following the introduction, formulation of multi-objective GP and corresponding weighting method have been discussed in section-2 and 3. The duality theory has been discussed in section-4 to find the optimal value of the objective function and the illustrative examples have been incorporated in section-5 to understand the problem. Finally, in section-6 some conclusions are drawn from the discussion.

## 2 Formulation of Multi-objective Geometric Programming

A multi-objective geometric programming problem can be defined as:

$$\text{Find } x = (x_1, x_2, \ldots x_n)^T$$

so as to

$$\min : f_{k0}(x) = \sum_{t=1}^{T_{k0}} g_k C_{k0t} \prod_{j=1}^{n} x_j^{a_{k0tj}}$$

$$k = 1, 2, \ldots, p \qquad (2.1)$$

subject to

$$f_i(x) = \sum_{t=1}^{T_i} C_{it} \prod_{j=1}^{n} x_j^{a_{itj}} \leq 1, i = 1, 2, \ldots, m \qquad (2.2)$$

$$x_j > 0, j = 1, 2, \ldots, n \qquad (2.3)$$

Where $C_{k0t}$ for all k and t are positive real numbers and $a_{itj}$ and $a_{k0tj}$ are real numbers for all $i, k, t, j$.
$g_k$ are continuous functions for all k.

$T_{k0}$ = number of terms present in the $k^{th}$ objective function.

$T_i$ = number of terms present in the $i^{th}$ constraint.

In the above multi-objective geometric program there are p number of minimization type objective function, m number of inequality type constraints and n number of strictly positive decision variables.

## 3 Weighting Method of Multi-objective Functions

The weighting method is the simplest multi-objective optimization which has been widely applied to find the non-inferior optimal solution of multi-objective function within the convex objective space.

If $f_{10}(x), f_{20}(x), \ldots, f_{p0}(x)$ are p objective functions for any vector $x = (x_1, x_2, \ldots x_n)^T$
then we can define weighting method for their optimal solution as defined below:

Let $\quad W = \{w : w \in R,^n w_k > 0, \sum_{k=1}^{n} w_k = 1\}$

be the set of non-negative weights. Using weighting method the above multi-objective function can be defined as:

$$Q(w) = \min_{x \in X} \sum_{k=1}^{p} w_k f_{k0}(x) \qquad (3.1)$$

subject to

$$f_i(x) \leq 1; \quad i = 1, 2, \ldots, m \qquad (3.2)$$

$$x_j > 0; \quad j = 1, 2, \ldots, n \qquad (3.3)$$

It must be made clear, however, that if the objective space of the original problem is non-convex, then the weighting method may not be capable of generating the efficient solutions on the non-convex part of the efficient frontier. It must also be noted that the optimal solution of a weighting problem should not be used as the best compromise solution, if the weights do not reflect the decision maker's preferences or if the decision maker does not accept the assumption of a linear utility function. For more details about the weighted method refer[13].

Based on the importance of the p number of objective functions defined in(2.1) the weights $w_1, w_2, \ldots, w_p$ are assigned to define a new min type objective function $Z(x)$ which can be defined as

$$\min_{x} : Z(x) = \sum_{k=1}^{p} w_k f_{k0}(x)$$

$$= \sum_{k=1}^{p} w_k \left(\sum_{t=1}^{T_{k0}} g_k C_{k0t} \prod_{j=1}^{n} x_j^{a_{k0tj}}\right)$$

$$= \sum_{k=1}^{p} \sum_{t=1}^{T_{k0}} w_k g_k C_{k0t} \prod_{j=1}^{n} x_j^{a_{k0tj}} \qquad (3.4)$$

$$x_j > 0, j = 1, 2, \ldots, n \qquad (3.5)$$

where

$$\sum_{k=1}^{p} w_k = 1, w_k > 0, k = 1, 2, \ldots, p \qquad (3.6)$$

## 4 Dual Form of GPP

The model given by (3.4), (3.5) and (3.6) is a conventional geometric programming problem and it can be solved directly by using primal based algorithm for non linear primal problem or dual programming [14]. Methods due to Rajgopal and Bricker [17], Beightler and Phillips[1] and Duffin et al.[6] projected in their analysis that the dual problem has the desirable features of being linearly constrained and having an objective function with structural properties with suitable solution.

According to Duffin et al.[6] the model given by (3.5) can be transformed to the corresponding dual geometric program as:

$$\max_{w} : \prod_{t=1}^{T_{k0}} \left(\frac{w_k g_k C_{k0t}}{w_{0t}}\right)^{w_{0t}} \prod_{i=1}^{m}\prod_{t=1}^{T_i} \left(\frac{w_{i0}}{w_{it}}\right)^{w_{it}} \lambda(w_{it})^{\lambda(w_{it})} \quad (4.1)$$

subject to

$$\sum_{t=1}^{T_{k0}} w_{0t} = 1$$

$$\sum_{k=1}^{p}\sum_{t=1}^{T_{k0}} a_{k0tj} w_{0t} + \sum_{i=1}^{m}\sum_{t=1}^{T_i} a_{itj} w_{it} = 0, j=1,2,\ldots,n$$

$$w_{it} \geq 0 \forall t,i$$

$$\sum_{k=1}^{p} w_k = 1, w_k > 0, k=1,2,\ldots,p$$

Since it is usually a dual problem then it can be solved using a method relating to the dual theory.

## 5 Numerical Examples

For illustration we consider the following examples.
Example:1 Find $x_1, x_2, x_3$ so as to
min : $f_1(x) = g x_1^{-1} x_2^{-1/2} x_3^{-1} + 20 x_1 x_3 + 20 x_1 x_2 x_3$ (5.1)
min : $f_2(x) = 40 x_1^{-1} x_2^{-1} x_3^{-1} + h x_1^{1/3} x_3^{3/4}$

subject to

$x_1^{-2} x_2^{-2} + 4 x_2^{1/2} x_3^{3/4} \leq 3$ (5.3)
$x_1, x_2, x_3 > 0$

In this example we have considered some cost coefficients in two objectives as continuous functions.

Introducing weights for the above objective functions a new objective function is formulated as:

$Z(x) = w_1(g x_1^{-1} x_2^{-1/2} x_3^{-1} + 20 x_1 x_3 + 20 x_1 x_2 x_3)$
$\quad + w_2(40 x_1^{-1} x_2^{-1} x_3^{-1} + h x_1^{1/3} x_3^{3/4})$ (5.4)

subject to

$(1/3) x_1^2 x_2^{-2} + (4/3) x_2^{1/2} x_3^{-1} \leq 1$ (5.5)
$x_1, x_2, x_3 > 0$

where

$w_1 + w_2 = 1, w_1, w_2 > 0$ (5.6)
$g(t) = 2t + 2$ (5.7)
$h(t) = t + 1$ (5.8)

This problem is having a certain degree of difficulty 3. The problem is solved via the dual programming Duffin [7]

The corresponding dual program is:

$$\max_{w} : V(w) = \left(\frac{g w_1}{w_{01}}\right)^{w_{01}} \left(\frac{20 w_1}{w_{02}}\right)^{w_{02}} \left(\frac{20 w_1}{w_{03}}\right)^{w_{03}}$$

$$\left(\frac{40 w_2}{w_{04}}\right)^{w_{04}} \left(\frac{h w_2}{w_{05}}\right)^{w_{05}} \left(\frac{\frac{1}{3}}{w_{11}}\right)^{w_{11}} \left(\frac{\frac{4}{3}}{w_{12}}\right)^{w_{12}}$$

$$(w_{11} + w_{12})^{(w_{11}+w_{12})} \quad (5.9)$$

subject to

$w_{01} + w_{02} + w_{03} + w_{04} + w_{05} = 1$
$-w_{01} + w_{02} + w_{03} - w_{04} + (1/3) w_{05} - 2 w_{11} = 0$
$-(1/2) w_{01} + w_{03} - w_{04} - 2 w_{11} + (1/2) w_{12} = 0$
$-w_{01} + w_{02} + w_{03} - w_{04} + (3/4) w_{05} - w_{12} = 0$
$w_1 + w_2 = 1$
$w_{01}, w_{02}, w_{03}, w_{04}, w_{05}, w_{11}, w_{12} \geq 0$
$w_1, w_2 > 0$

By considering different values of $w_1, w_2, g, h$ and the dual variables, corresponding maximum value of dual objective is given in the following table.

Table-1(a)
Dual Solution [g=40,h=20]

| $w_1$ | $w_2$ | $w_{01}$ | $w_{02}$ | $w_{03}$ |
|---|---|---|---|---|
| .1 | .9 | .3538166E-01 | .4491953E-01 | .1629705 |
| .2 | .8 | .6460256E-01 | .9304868E-01 | .2224467 |
| .3 | .7 | .8981589E-01 | .1386930 | .2615291 |
| .4 | .6 | .1117080 | .1822681 | .2912977 |
| .5 | .5 | .1308674 | .2238612 | .3157730 |

Table-1(b)
Dual Solution[g=40,h=20

| $w_{04}$ | $w_{05}$ | $w_{11}$ | $w_{12}$ | Z |
|---|---|---|---|---|
| .1671291 | .5895484 | .1009222 | .4474897 | 51.40669 |
| .1671291 | .4527729 | .1173440 | .4233434 | 58.67673 |
| .1526150 | .3573470 | .1384534 | .4258014 | 64.87778 |
| .1325447 | .2821815 | .1616868 | .4409492 | 70.51816 |
| .1101877 | .2193107 | .1858413 | .4630621 | 75.81600 |

Table-2(a)
Dual Solution [g=42,h=21]

| $w_1$ | $w_2$ | $w_{01}$ | $w_{02}$ | $w_{03}$ |
|---|---|---|---|---|
| .1 | .9 | .3595181E-01 | .4399913E-01 | .1570954 |
| .2 | .8 | .6595986E-01 | .9141579E-01 | .2162172 |
| .3 | .7 | .9206252E-01 | ..1364239 | .2553973 |
| .4 | .6 | .1148757 | .1794227 | .2853658 |
| .5 | .5 | .1349398 | .2205432 | .3100875 |

Table-2(b)
Dual Solution [g=42,h=21]

| $w_{04}$ | $w_{05}$ | $w_{11}$ | $w_{12}$ | Z |
|---|---|---|---|---|
| .1630853 | .5998684 | .1010068 | .4519587 | 53.01235 |
| .1633865 | .4630206 | .1163134 | .4255522 | 60.19380 |
| .1495228 | .3665935 | .1362168 | .4251809 | 66.32737 |
| .1301272 | .2902086 | .1582609 | .4374420 | 71.90609 |
| .1083816 | .2260479 | .1813293 | .4568453 | 77.14276 |

Table-3(a)
Dual Solution [g=44,h=22]

| $w_1$ | $w_2$ | $w_{01}$ | $w_{02}$ | $w_{03}$ |
|---|---|---|---|---|
| .1 | .9 | .364869E-01 | .431282E-01 | .1515729 |
| .2 | .8 | .672419E-01 | .898600E-01 | .2102984 |
| .3 | .7 | .941988E-01 | .1342586 | .2495379 |
| .4 | .6 | .1179082 | .1766990 | .2796733 |
| .5 | .5 | .1388617 | .2173522 | .3046089 |

Table-3(b)
Dual Solution [g=44,h=22]

| $w_{04}$ | $w_{05}$ | $w_{11}$ | $w_{12}$ | Z |
|---|---|---|---|---|
| .1592421 | .6095698 | .1010810 | .4561494 | 54.61683 |
| .1598354 | .47276742 | .1153346 | .4276542 | 61.70927 |
| .1465658 | .3754388 | .1340890 | .4246109 | 67.77508 |
| .1278009 | .2979186 | .1549847 | .4341022 | 73.29226 |
| .1066352 | .2325419 | .1769891 | .4508707 | 78.46815 |

Using primal dual relationship the corresponding primal solution are given in the following table.

Table-4
Primal Solution[g=40,h=20]

| $w_1$ | $w_2$ | $x_1$ | $x_2$ | $x_3$ | Z |
|---|---|---|---|---|---|
| 0.1 | 0.9 | 0.3709596 | 3.628046 | 3.112426 | 51.40669 |
| 0.2 | 0.8 | 0.5184022 | 2.390645 | 2.632994 | 58.67673 |
| 0.3 | 0.7 | 0.6181019 | 1.885667 | 2.426272 | 64.87778 |
| 0.4 | 0.6 | 0.6974348 | 1.598181 | 2.303659 | 70.51816 |
| 0.5 | 0.5 | 0.7648254 | 1.410574 | 2.219104 | 75.81600 |

Table-5
Primal Solution[g=42,h=21]

| $w_1$ | $w_2$ | $x_1$ | $x_2$ | $x_3$ | Z |
|---|---|---|---|---|---|
| 0.1 | 0.9 | 0.3783509 | 3.570412 | 3.082458 | 53.01235 |
| 0.2 | 0.8 | 0.5268679 | 2.365203 | 2.611029 | 60.19380 |
| 0.3 | 0.7 | 0.6260859 | 1.872084 | 2.408787 | 66.32737 |
| 0.4 | 0.6 | 0.7042765 | 1.590465 | 2.289866 | 71.90609 |
| 0.5 | 0.5 | 0.7703444 | 1.406016 | 2.208535 | 77.14276 |

Table-6
Primal Solution [g=44, h=22]

| $w_1$ | $w_2$ | $x_1$ | $x_2$ | $x_3$ | Z |
|---|---|---|---|---|---|
| 0.1 | 0.9 | 0.3857122 | 3.514460 | 3.053485 | 54.61683 |
| 0.2 | 0.8 | 0.5352863 | 2.340286 | 2.589830 | 61.70927 |
| 0.3 | 0.7 | 0.6340715 | 1.858634 | 2.391791 | 67.77508 |
| 0.4 | 0.6 | 0.7111610 | 1.582766 | 2.276326 | 73.29226 |
| 0.5 | 0.5 | 0.7759240 | 1.401452 | 2.198055 | 78.46815 |

**Example:2**

Find $x_1, x_2, x_3, x_4$ so as to

min : $f_1(x) = gx_1x_2^2 x_3^{-1} + 2x_1^{-1} x_2^{-3}x_4 + 10x_1x_3$  (5.10)

min : $f_2(x) = x_1^{-2} x_2^{-1} x_3 x_4^{-1} + hx_1^{-3} x_2^2 x_3^{-2}$  (5.11)

subject to

$$3x_1^{-1}x_3x_4^{-2} + 4x_3x_4 \leq 1$$  (5.12)
$$kx_1x_2 \leq 1$$  (5.13)
$$x_1, x_2, x_3, x_4 > 0$$  (5.14)

In this example we have considered some cost coefficients in two objectives are continuous functions and one constraint coefficient is continuous function. Using the weights the above objective function can be reduced to the new objective function as:

$Z(x) = w_1(gx_1x_2^2 x_3^{-1} + 2x_1^{-1}x_2^{-3} + 10x_1x_3)$
$\quad\quad + w_2(x_1^2x_2^{-1}x_3x_4^{-1} + hx_1^{-3}x_2^2 x_3^{-2})$  (5.15)

subject to

$$3x_1^{-1}x_3x_4^{-2} + 4x_3x_4 \leq 1$$  (5.16)
$$kx_1x_2 \leq 1$$  (5.17)
$$x_1, x_2, x_3 > 0$$  (5.18)

where

$$g(t) = t^2$$  (5.19)
$$h(t) = t + 1$$  (5.20)
$$k(t) = 2t + 3$$  (5.21)
$$w_1 + w_2 = 1, w_1, w_2 > 0$$  (5.22)

In this problem the degree of difficulty is 3 and it can be solved by using duality theory as given by

| $w_1$ | $w_2$ | $w_{01}$ | $w_{02}$ | $w_{03}$ | $w_{04}$ |
|---|---|---|---|---|---|
| .1 | .9 | .2583442 E-02 | .7428173 | .1483374 E-02 | .2118867 E-01 |
| .2 | .8 | .3764835 E-02 | .7584524 | .1511017 E-02 | .8334759 E-02 |
| .3 | .7 | .4753309 E-02 | .7629485 | .1538947 E-02 | .4322214 E-02 |
| .4 | .6 | .5718845 E-02 | .7646817 | .1566427 E-02 | .2491526 E-02 |
| .5 | .5 | .6754852 E-02 | .7653131 | .1594399 E-02 | .1493452 E-02 |

$$\max_{w}: V(w) = \left(\frac{gw_1}{w_{01}}\right)^{w_{01}} \left(\frac{2w_1}{w_{02}}\right)^{w_{02}} \left(\frac{10w_1}{w_{03}}\right)^{w_{03}} \left(\frac{w_2}{w_{04}}\right)^{w_{04}} \left(\frac{hw_2}{w_{05}}\right)^{w_{05}} \left(\frac{3}{w_{11}}\right)^{w_{11}} \left(\frac{4}{w_{12}}\right)^{w_{12}} k^{w_{21}} (w_{11}+w_{12})^{(w_{11}+w_{12})}$$

subject to

$$w_{01} + w_{02} + w_{03} + w_{04} + w_{05} = 1$$
$$w_{01} - w_{02} + w_{03} + 2w_{04} - 3w_{05} - w_{11} + w_{21} = 0$$
$$2w_{01} - 3w_{02} - w_{03} + 2w_{04} + w_{21} = 0$$
$$-w_{01} + w_{03} + w_{04} - 2w_{05} + w_{11} + w_{12} = 0$$
$$w_{02} - w_{04} - 2w_{11} + w_{12} = 0$$
$$w_{01}, w_{02}, w_{03}, w_{04}, w_{05}, w_{11}, w_{12}, w_{21} \geq 0$$
$$w_1 + w_2 = 1$$
$$w_1, w_2 > 0$$

For different values of $w_1, w_2, g, h$ and the dual variables; the corresponding maximum values of dual objectives are obtained as given in the tables

Table-1(a)

Dual Solution [g=1,h=2,k=5]

| $w_1$ | $w_2$ | $w_{01}$ | $w_{02}$ | $w_{03}$ | $w_{04}$ |
|---|---|---|---|---|---|
| .1 | .9 | .1883184 E-02 | .7066979 | .3912119 E-02 | .4807176 E-01 |
| .2 | .8 | .2875885 E-02 | .7404825 | .3957753 E-02 | .2103163 E-01 |
| .3 | .7 | .3709217 E-02 | .7521911 | .3999773 E-02 | .1141363 E-01 |
| .4 | .6 | .4515186 E-02 | .7576144 | .4050324 E-02 | .6740309 E-02 |
| .5 | .5 | .5731889 E-02 | .7604136 | .4108534 E-02 | .4099518 E-02 |

Table-1(b)

Dual Solution [g=1,h=2,k=5]

| $w_{05}$ | $w_{11}$ | $w_{12}$ | $w_{21}$ | Z |
|---|---|---|---|---|
| .2394350 | .3624652 | .6630413 E-01 | 1.685529 | 23.30086 |
| .2316522 | .3868806 | .5431305 E-01 | 1.773423 | 37.49259 |
| .2286863 | .3954820 | .5018649 E-01 | 1.803196 | 49.25933 |
| .2270798 | .3995861 | .4829802 | 1.816394 | 59.15666 |
| .2260064 | .4018303 | .4734644 E-01 | 1.822584 | 67.31155 |

Table-2(a)

Dual Solution [g=4,h=3,k=7]

Table-2(b)

Dual Solution [g=4,h=3,k=7]

| $w_{05}$ | $w_{11}$ | $w_{12}$ | $w_{21}$ | Z |
|---|---|---|---|---|
| .2319272 | .3884648 | .5530093 E-01 | 1.780619 | 46.15296 |
| .2279370 | .3999702 | .4982283 E-01 | 1.820288 | 75.76130 |
| .2264370 | .4034642 | .4830204 E-01 | 1.830787 | 100.2160 |
| .2255415 | .4049780 | .4776588 E-01 | 1.834016 | 120.7471 |
| .2248442 | .4057250 | .4763040 E-01 | 1.834235 | 137.6543 |

Table-3(a)

Dual Solution [g=9,h=4,k=9]

| $w_1$ | $w_2$ | $w_{01}$ | $w_{02}$ | $w_{03}$ | $w_{04}$ |
|---|---|---|---|---|---|
| .1 | .9 | .2556200 E-02 | .7569191 | .7197187 E-03 | .1069413 E-01 |
| .2 | .8 | .3643589 E-02 | .7647673 | .7388184 E-03 | .4008480 E-02 |
| .3 | .7 | .4564563 E-02 | .7667072 | .7551118 E-03 | .2044494 E-02 |
| .4 | .6 | .5471832 E-02 | .7672706 | .7699552 E-03 | .1169431 E-02 |
| .5 | .5 | .6450544 E-02 | .7672784 | .7844841 E-03 | .6980240 E-03 |

Table-3(b)

Dual Solution [g=9,h=4,k=9]

| $w_{05}$ | $w_{11}$ | $w_{12}$ | $w_{21}$ | Z |
|---|---|---|---|---|
| .2291109 | .3985297 | .5083442 E-01 | 1.818117 | 77.71193 |
| .2268418 | .4044462 | .4813366 E-01 | 1.837340 | 128.4884 |
| .2259287 | .4060950 | .4752731 E-01 | 1.841180 | 170.3272 |
| .2253182 | .4067567 | .4741216 E-01 | 1.841401 | 205.4137 |
| .2247886 | .4070418 | .4750335 E-01 | 1.840055 | 234.2872 |

The corresponding primal solutions are given in the following tables:

Table-4

Primal Solution[g=1,h=2,k=5]

| $w_1$ | $w_2$ | $x_1$ | $x_2$ | $x_3$ | $x_4$ | Z |
|---|---|---|---|---|---|---|
| .1 | .9 | 1.158065 | 0.1727019 | 0.7871444 E-01 | 0.4911339 | 23.30086 |
| .2 | .8 | 1.102102 | 0.1814714 | 0.6732011 E-01 | 0.4571373 | 37.49259 |
| .3 | .7 | 1.048456 | 0.1907567 | 0.6264038 E-01 | 0.4494227 | 49.25933 |
| .4 | .6 | 0.9986112 | 0.2002781 | 0.5998444 E-01 | 0.4494268 | 59.15666 |
| .5 | .5 | 0.9507755 | 0.2103546 | 0.5817415 E-01 | 0.4529743 | 67.31155 |

Table-5

Primal Solution[g=4,h=3,k=7]

Table-6

Primal Solution [g=9,h=4,k=9]

| $w_1$ | $w_2$ | $x_1$ | $x_2$ | $x_3$ | $x_4$ | Z |
|---|---|---|---|---|---|---|
| .1 | .9 | 0.9275071 | 0.1197954 | 0.6030440 E-01 | 0.4689700 | 77.71193 |
| .2 | .8 | 0.8441800 | 0.1316202 | 0.5622785 E-01 | 0.4728634 | 128.4884 |
| .3 | .7 | 0.7876750 | 0.1410621 | 0.5443038 E-01 | 0.4812168 | 170.3272 |
| .4 | .6 | 0.7425651 | 0.1496315 | 0.5324933 E-01 | 0.4901075 | 205.4137 |
| .5 | .5 | 0.7026785 | 0.1581251 | 0.5231420 E-01 | 0.4994140 | 234.2872 |

## 6 Conclusions

By using weighted method we can solve a multi-objective GPP as a vector-minimum problem. A vector-maximum problem can be transformed as a vector-minimization problem. If any of the objective function and/or constraint does not satisfy the property of a posynomial after the transformation, then we use any of the general purpose non-linear programming algorithms to solve the problem. We can also use this technique to solve a multi-objective signomial geometric programming problem. However, if a GPP has either a higher degree of difficulty or a negative degree of difficulty, then we can use any of the general purpose non-linear programming algorithm instead of a GP algorithm.

## 7 Acknowledgements

| $w_1$ | $w_2$ | $x_1$ | $x_2$ | $x_3$ | $x_4$ | Z |
|---|---|---|---|---|---|---|
| .1 | .9 | 1.031732 | 0.1384634 | 0.6635765 E-01 | 0.4694868 | 46.15296 |
| .2 | .8 | 0.9530635 | 0.1498926 | 0.6005816 E-01 | 0.4610817 | 75.76130 |
| .3 | .7 | 0.8940238 | 0.1597912 | 0.5750384 E-01 | 0.4648240 | 100.2160 |
| .4 | .6 | 0.8449595 | 0.1690698 | 0.5596250 E-01 | 0.4713043 | 120.7471 |
| .5 | .5 | 0.8006811 | 0.1784195 | 0.5482305 E-01 | 0.4790883 | 137.6543 |

**Dr. A.K. Ojha:** Dr. A.K. Ojha received a Ph.D(mathematics) from Utkal University in 1997. Currently he is an Asst. Prof. in Mathematics at I.I.T. Bhubaneswar, India. He is performing research in Neural Network, Geometric Programming, Genetical Algorithm, and Particle Swarm Optimization. He has served more than 27 years in different Govt. colleges in the state of Orissa. He has published 22 research papers in different journals and 7 books for degree students such as: Fortran 77 Programming, A text book of modern algebra, Fundamentals of Numerical Analysis etc.

**A.K. Das:** Mr. A.K. Das received a M.Sc. (Mathematics) from Ravenshaw(Auto) College,Cuttack,Orissa in 1997. Currently he is a lecturer in Mathematics at Bhadrak College,Bhadrak,Orissa, India. He is performing research works in Geometric Programming. He has served more than 10 years in different colleges in the state of Orissa. He has published 2 research papers in different journals.